\documentclass[english,showpacs]{revtex4}
\usepackage[T1]{fontenc}
\usepackage[latin9]{inputenc}
\usepackage{amsthm}
\usepackage{amsmath}
\usepackage{graphicx}
\usepackage{esint}

\usepackage{babel}

\newcommand{\BrSbsSRpi}{18}  \newcommand{\BrSbsPMpi}{12}  \newcommand{\BrSbsLFpi}{5.5}
\newcommand{\BrVbsSRpi}{7}  \newcommand{\BrVbsPMpi}{9.4}  \newcommand{\BrVbsLFpi}{6.2}

\newcommand{\BrSbsSRrho}{7.6}  \newcommand{\BrSbsPMrho}{5.4}  \newcommand{\BrSbsLFrho}{3.1}
\newcommand{\BrVbsSRrho}{21}  \newcommand{\BrVbsPMrho}{22}  \newcommand{\BrVbsLFrho}{15}

\newcommand{\BrSbsSRpp}{6.1}  \newcommand{\BrSbsPMpp}{4.3}  \newcommand{\BrSbsLFpp}{2.4}
\newcommand{\BrVbsSRpp}{13}  \newcommand{\BrVbsPMpp}{14}  \newcommand{\BrVbsLFpp}{8.3}

\newcommand{\BrSbsSRppp}{0.096}  \newcommand{\BrSbsPMppp}{0.068}  \newcommand{\BrSbsLFppp}{0.039}
\newcommand{\BrVbsSRppp}{0.23}  \newcommand{\BrVbsPMppp}{0.24}  \newcommand{\BrVbsLFppp}{0.16}

\newcommand{\BrSbsSRpppp}{0.0064}  \newcommand{\BrSbsPMpppp}{0.0045}  \newcommand{\BrSbsLFpppp}{0.0026}
\newcommand{\BrVbsSRpppp}{0.015}  \newcommand{\BrVbsPMpppp}{0.016}  \newcommand{\BrVbsLFpppp}{0.01}

\newcommand{\BrSbsSRqq}{19}  \newcommand{\BrSbsPMqq}{13}  \newcommand{\BrSbsLFqq}{6.5}
\newcommand{\BrVbsSRqq}{20}  \newcommand{\BrVbsPMqq}{23}  \newcommand{\BrVbsLFqq}{13}

\begin{document}

\title{Light hadron production in $B_{c}\to B_{s}^{(*)}+X$ decays}

\author{A.K. Likhoded}

\email{Anatolii.Likhoded@ihep.ru}

\affiliation{Institute of High Energy Physics, Protvino, Russia}

\author{A.V. Luchinsky}

\email{Alexey.Luchinsky@ihep.ru}

\affiliation{Institute of High Energy Physics, Protvino, Russia}
\begin{abstract}
The article is devoted to $B_{c}\to B_{s}^{(*)}+n\pi$ decays with
$n=1$, $2$, $3$, 4. In the framework of factorization theorem the
branching fractions of these processes can be written as convolution
of hard part, describing $B_{c}\to B_{s}^{(*)}W$ vertices, and spectral
functions, that correspond to transition of virtual $W$-boson into
a final $\pi$-meson system. These functions were obtained from the
fit of experimental data on $\tau$-lepton decay and electron-positron
annihilation. Using different sets of $B_{c}\to B_{s}^{(*)}W$ decay
form-factors we present branching fractions and distributions over
the invariant mass of the final $\pi$-meson system.
\end{abstract}


\maketitle

\section{Introduction}

$B_{c}$-meson is the heaviest of the particles stable with respect to strong and electromagnetic interaction. The decays of the ground state of $\left(\bar{b}c\right)$-system can be caused only by weak interaction, with either $c$-quark decays, $b$-quark decays or annihilation allowed. According to \cite{Gershtein1995}, the ratios of these processes are $\sim45\%$, 37\% and 18\% respectively. Up to now only two decay modes, caused by $b$-quark decay, are observed: $B_{c}\to J/\psi\pi$ and semileptomic decay $B_{c}\to J/\psi\ell\nu$. With the help of the former mode $B_{c}$-meson mass was determined with pretty good accuracy: $m_{B_{c}}=6275.6\pm2.9\pm2.5$ MeV (CDF collaboration \cite{Aaltonen2007gv}) and $m_{B_{c}}=6300\pm 14\pm 5$ MeV (D0 collaboration \cite{Abazov2008}).
The latter decay mode gives the opportunity to measure $B_{c}$-meson
life time: $\tau_{B_{c}}=0.448_{-0.036}^{+0.038}\pm0.032$ ps (D0
collaboration \cite{Abazov2008rb}) and $\tau_{B_{c}}=0.475_{-0.049}^{+0.053}\pm0.018$
ps (CDF collaboration \cite{Abe1998,Abulencia2006zu}). Both the mass
of $B_{c}$-meson and its lifetime are in good agreement with theoretical
predictions based on potential quark models and QCD sum rules \cite{Kiselev2000nf,Kiselev2000pp,Kiselev1999sc,Kiselev2002vz,Huang2007kb}.

According to perturbative QCD estimates \cite{Gershtein1997qy}, the
cross section of $B_{c}$-meson production in hadronic experiments
is about $10^{-3}$ of $B$-meson production cross section. As a result,
one can expect about $10^{9}$ $B_{c}$-mesons at LHC collider luminosity
$\sim1\,\mathrm{fb}^{-1}$, so detailed investigation of ground $B_{c}$-meson
and excited states of $\left(\bar{b}c\right)$-family (there are 16
narrow states below $BD$-pair production threshold) would be possible.

In our previous work \cite{Likhoded2009} we considered $B_{c}$-meson
decays with $b$-quark as a spectator, namely the reactions $B_{c}\to J/\psi+X$,
where $X$ stands for lepton pairs, light quarks $u\bar{d}$ or $\pi$-meson
system $n\pi$ for $1\le n\le4$. In the framework of factorization
theorem the amplitude of this process can be written as a convolution
of $B_{c}\to J/\psi W$-decay width and spectral functions that describe
the transition of virtual $W$-boson into final $\pi$-mesons state.
These spectral functions do not depend on $W$-boson production mechanism,
so they can be determined from experimental and theoretical analysis
of other reactions, for example $\tau$-lepton decay $\tau\to\nu_{\tau}+X$
or electron-positron annihilation $e^{+}e^{-}\to X$. In the current
work we consider induced by $c$-quark weak decay reactions $B_{c}\to B_{s}^{(*)}+X$.
In contrast to $B_{c}\to J/\psi+X$ decays, these processes are Cabbibo-allowed,
so their branching fractions are about an order higher than the branching
fractions of $B_{c}\to J/\psi+X$ decays. Form theoretical point of
view these branching fractions are determined, from one side, by effective
hamiltonian of weak $c\to s$ decay, where all higher-order QCD corrections
are taken into account, and, on the other hand, by form-factors of
$B_{c}\to B_{s}^{(*)}$ transitions. These from-factors can be obtained
in different models, for example QCD sum rules \cite{Kiselev1992tx,Gershtein1995,Kiselev2000nf,Kiselev2000pp,Kiselev1999sc},
potential quark models \cite{Gershtein1995,Colangelo1999zn,Kiselev2002vz,Hernandez2006gt},
light-front approach \cite{Anisimov1998xv,Anisimov1998uk,Choi2009,Choi2009ai,Wang2009mi}.
Calculated with these form factors widths of $B_{c}\to B_{s}^{(*)}+X$
decays differ from one another and, since considered here decays are
dominant, this leads to different predictions for $B_{c}$-meson lifetime.
As we have mentioned above, the lifetime of $B_{c}$-meson is known
experimentally with pretty good accuracy, so this difference can be
used to select physical set of form-factors.

In the next section we present analytical expressions for transferred
momentum distributions of branching fractions of the decays $B_{c}\to B_{c}^{(*)}+X$,
where $X$ is lepton pair $\ell\nu$ or light meson system. These
distributions are expressed through spectral functions of the final
state $X$. In section III explicit expressions of spectral functions,
obtained from analysis of $\tau$-lepton decay $\tau\to\nu_{\tau}+X$
and electron-positron annihilation $e^{+}e^{-}\to X$, are given.
Using these expressions we give predictions for branching fractions
of the decays $B_{c}\to B_{s}^{(*)}+n\pi$ and distributions over
the squared momentum of light mesons system. In the last section of
our paper we give the conclusion.

\section{Analytical Results}

\begin{figure}
\begin{centering}
\includegraphics[scale=0.6]{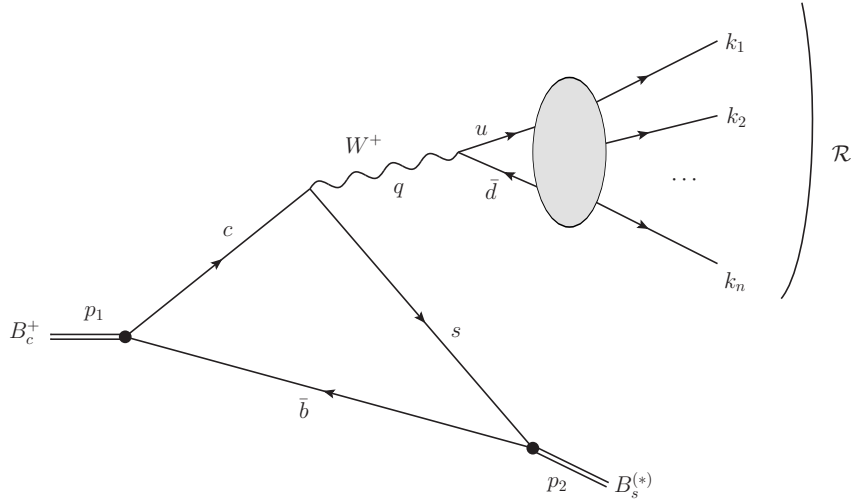}
\par\end{centering}

\caption{\label{fig:diag}Typical diagram for $B_{c}\to B_{s}^{(*)}+n\pi$
decay}

\end{figure}

Decays $B_{c}\to B_{s}^{(*)}+n\pi$ are caused mainly by weak $c$-quark
decay $c\to sW^{+}\to u\bar{d}$ (see diagram shown in fig.\ref{fig:diag}),
while $b$-quark is a spectator. Effective Lagrangian of this process
has the form

\begin{eqnarray*}
\mathcal{H}_{\mbox{eff}} & = & \frac{G_{F}}{2\sqrt{2}}V_{cs}V_{ud}^{*}\left[C_{+}\left(\mu\right)O_{+}+C_{-}\left(\mu\right)O_{-}\right],\end{eqnarray*}
 where $G_{F}$ is Fermi coupling constant, $V_{ij}$ are CKM mixing
matrix elements and $C_{\pm}(\mu)$ are Wilson coefficients, that
describes higher-order QCD corrections. The operators $O_{\pm}$are
defined according to

\begin{eqnarray*}
O_{\pm} & = & \left(\bar{d}_{i}u_{j}\right)_{V-A}\left(\bar{c}_{i}b_{j}\right)_{V-A}\pm\left(\bar{d}_{j}u_{i}\right)_{V-A}\left(\bar{c}_{i}b_{j}\right)_{V-A},\end{eqnarray*}
 where $i$, $j$ are colour indexes of quarks and $\left(\bar{q}_{1}q_{2}\right)_{V-A}=\bar{q}_{1}\gamma_{\mu}(1-\gamma_{5})q_{2}$.
Since in the considered here decays the light-quark pair is in colour-singlet
state, the amplitude of these processes should be proportional to
factor

\begin{eqnarray*}
a_{1}(\mu) & = & \frac{1}{2N_{c}}\left[\left(N_{c}-1\right)C_{+}\left(\mu\right)+\left(N_{c}+1\right)C_{-}\left(\mu\right)\right].\end{eqnarray*}
 If QCD corrections are neglected, this fuction is equal to $a_{1}\left(\mu\right)=1$.
Due to higher-order logarithmic corrections the dependence of this
coefficient on renormalisation scale $\mu$ appears \cite{Buchalla1996},
and on $\mu\sim m_{b}$ it is equal to \cite{Kiselev2002vz}

\begin{eqnarray*}
a_{1}\left(m_{b}\right) & = & 1.2.\end{eqnarray*}

The matrix element of the decay $B_{c}\to B_{c}^{(*)}+\mathcal{R}$,
where $\mathcal{R}$ is some set of light hadrons, has the form

\begin{eqnarray*}
\mathcal{M}\left[B_{c}^{+}\to B_{s}^{(*)}W^{+}\to B_{s}^{(*)}+\mathcal{R}\right] & = & \frac{G_{F}V_{cs}}{\sqrt{2}}a_{1}\mathcal{H}_{\mu}\epsilon_{\mu}^{\mathcal{R}},\end{eqnarray*}
 where $\epsilon^{\mathcal{R}}$ is the effective polarization vector
of light hadron system $\mathcal{R}$, and $\mathcal{H}$ vertex is

\begin{eqnarray*}
\mathcal{H}_{\mu} & = & f_{+}\left(q^{2}\right)p_{\mu}+f_{-}\left(q^{2}\right)q_{\mu}\end{eqnarray*}
 for $B_{c}$-meson decay to pseudo-scalar $B_{s}$-meson and

\begin{eqnarray*}
\mathcal{H}_{\mu} & = & \epsilon_{\mu}F_{0}^{A}\left(q^{2}\right)+\left(\epsilon p_{1}\right)p_{\mu}F_{+}^{A}\left(q^{2}\right)+\left(\epsilon p_{1}\right)q_{\mu}F_{-}^{A}\left(q^{2}\right)-ie^{\mu\nu\alpha\beta}\epsilon_{\nu}p_{\alpha}q_{\beta}F_{V}\left(q^{2}\right)\end{eqnarray*}
 for $B_{c}$-meson decay to vector $B_{s}^{*}$-meson. In these expressions
$p_{1,2}$ are the momenta of $B_{c}$- and $B_{s}^{(*)}$-mesons
respectively, $p=p_{1}+p_{2}$, $q=p_{1}-p_{2}$ is the momentum of
virtual $W$-boson, and $f_{\pm}\left(q^{2}\right)$, $F_{0,\pm}^{A}\left(q^{2}\right)$
and $F_{V}\left(q^{2}\right)$ are the form-factors of $B_{c}\to B_{s}^{(*)}W^{*}$
transition.

From presented above amplitudes it is easy to calculate the widths
of the $B_{c}\to B_{s}^{(*)}+\mathcal{R}$ decays:

\begin{eqnarray*}
d\Gamma\left[B_{c}\to B_{s}^{(*)}+\mathcal{R}\right] & = & \frac{1}{2M}\frac{G_{F}^{2}V_{cs}^{2}}{2}a_{1}^{2}\mathcal{H}_{\mu}\mathcal{H}_{\nu}^{*}\epsilon_{\mathcal{R}}^{\mu}\epsilon_{\mathcal{R}}^{*\nu}d\Phi\left(B_{c}\to B_{s}^{(*)}+\mathcal{R}\right)\end{eqnarray*}
 where $M=M_{B_{c}}$is the mass of initial $B_{c}$-meson, and Lorentz-invariant
phase space is defined according to

\begin{eqnarray*}
d\Phi\left(Q\to k_{1}\dots k_{n}\right) & = & \left(2\pi\right)^{4}\delta^{4}\left(Q-\sum_{i}k_{i}\right)\prod_{i}\frac{d^{3}k_{i}}{\left(2\pi\right)^{3}2E_{i}}.\end{eqnarray*}
 The following recurrence relation holds for this expression:

\begin{eqnarray*}
d\Phi\left(B_{c}\to B_{s}^{(*)}\mathcal{R}\right) & = & \frac{dq^{2}}{2\pi}d\Phi\left(B_{c}\to B_{s}^{(*)}W^{*}\right)d\Phi\left(W^{*}\to\mathcal{R}\right),\end{eqnarray*}
 Using it one can perform the integration over the phase space of
light hadrons system $\mathcal{R}$:

\begin{eqnarray*}
\frac{1}{2\pi}\int d\Phi\left(W^{*}\to\mathcal{R}\right)\epsilon_{\mu}^{\mathcal{R}}\epsilon_{\nu}^{*\mathcal{R}} & = & \left(q_{\mu}q_{\nu}-q^{2}g_{\mu\nu}\right)\varrho_{T}^{\mathcal{R}}\left(q^{2}\right)+q_{\mu}q_{\nu}\rho_{L}^{\mathcal{R}}\left(q^{2}\right).\end{eqnarray*}
 In the framework of factorization theorem introduced here spectral
functions $\rho_{T,L}^{\mathcal{R}}\left(q^{2}\right)$ are universal,
so they can be determined from theoretical and experimental analysis
of other reactions, for example $\tau$-lepton decay $\tau\to\nu_{\tau}+\mathcal{R}$
\cite{Schael2005} or electron-positron annihilation $e^{+}e^{-}\to\mathcal{R}$.
Explicit expressions of these spectral functions for different final
states $\mathcal{R}$ are presented in the next section.

From presented above matrix elements it is easy to obtain squared
transferred momentum distributions for the considered in our article
decays. In the case of pseudo-scalar $B_{s}$-meson we have

\begin{eqnarray*}
\frac{d\Gamma\left(B_{c}\to B_{s}+\mathcal{R}\right)}{dq^{2}} & = & \frac{G_{F}^{2}V_{cs}^{2}a_{1}^{2}}{32\pi M}\beta\left\{ \left|f_{+}\right|^{2}\left[M^{4}\beta^{2}\rho_{T}^{\mathcal{R}}+\left(M^{2}-m^{2}\right)^{2}\rho_{L}^{\mathcal{R}}\right]+\right.\\
 &  & \left.+\left|f_{-}^{2}\right|^{2}q^{4}\rho_{L}^{\mathcal{R}}+2\mathrm{Re}\left[f_{+}f_{-}^{*}\right]q^{2}\left(M^{2}-m^{2}\right)\rho_{T}^{\mathcal{R}}\right\} ,\end{eqnarray*}
 where $M$ and $m$ are the masses of $B_{c}$- and $B_{s}^{(*)}$-mesons
respectively, and

\begin{eqnarray*}
\beta & = & \sqrt{\frac{\left(M-m\right)^{2}-q^{2}}{M^{2}}}\sqrt{\frac{\left(M+m\right)^{2}-q^{2}}{M^{2}}}\end{eqnarray*}
 is the velocity of $B_{s}$-meson in $B_{c}$-meson rest frame. In
the case of vector $B_{s}$-meson in the final state the distribution
has the form

\begin{eqnarray*}
\frac{d\Gamma\left(B_{c}\to B_{s}^{*}+\mathcal{R}\right)}{dq^{2}} & = & \frac{128G_{F}^{2}V_{cs}^{2}a_{1}^{2}}{128\pi}\frac{M^{3}}{m^{2}}\beta\left\{ \rho_{T}^{\mathcal{R}}\left[\left(12\frac{q^{2}m^{2}}{M^{4}}+\beta^{2}\right)\left|F_{0}^{A}\right|^{2}+M^{4}\beta^{4}\left|F_{+}^{A}\right|^{2}+\right.\right.\\
 & + & \left.8m^{2}q^{2}\beta^{2}\left|F_{V}\right|^{2}+2\left(M^{2}-m^{2}-q^{2}\right)\beta^{2}\mathrm{Re}\left(F_{0}^{A}F_{+}^{A*}\right)\right]+\\
 & + & \rho_{L}^{\mathcal{R}}\beta^{2}\left[\left|F_{0}^{A}\right|^{2}+\left(M^{2}-m^{2}\right)^{2}\left|F_{+}^{A}\right|^{2}+q^{4}\left|F_{-}^{A}\right|^{2}+\right.\\
 & + & 2q^{2}\left(M^{2}-m^{2}\right)\mathrm{Re}\left(F_{0}^{A}F_{+}^{A*}\right)+2q^{2}\mathrm{Re}\left(F_{0}^{A}F_{-}^{A*}\right)+\\
 & + & \left.\left.2q^{2}\left(M^{2}-m^{2}\right)\mathrm{Re}\left(F_{+}^{A}F_{-}^{A*}\right)\right]\right\} .\end{eqnarray*}
 One can easily see, the $F_{V}\left(q^{2}\right)$ from-factor enter
in this expression only quadratically, so it is not possible to determine
its relative phase from transferred momentum distributions.

There are several; theoretical models, that predict $B_{c}\to B_{s}+W$
decay form-factors: QCD sum rules \cite{Kiselev2002vz}, potential
quark models \cite{Kiselev2002vz}, covariant light-front models \cite{Wang2009xt},
etc. These from-factors can be parametrised in different forms, for
example
\begin{enumerate}
\item monopole expression\begin{eqnarray*}
F_{i}\left(q^{2}\right) & = & \frac{F\left(0\right)}{1-q^{2}/m_{\mathrm{pole}}^{2}},\end{eqnarray*}

\item Isgur-Wise function $\xi(w)$
\item exponential parametrization, suitable for potential models, where
form-factors are determined by integral of initial and final quarkonia
wave-functions, that have Gaussian from:
\end{enumerate}
\begin{eqnarray}
F_{i}\left(q^{2}\right) & = & F\left(0\right)\exp\left\{ c_{1}q^{2}+c_{2}q^{4}\right\} .\label{eq:exp}\end{eqnarray}
 All parametrizations are almost equivalent and in our article we
use this the exponential parametrization (\ref{eq:exp}) for all sets
of form-factors. Numerical values of the parameters $F\left(0\right)$,
$c_{1,2}$ for different from-factors stets are given in table.\ref{tab:FFS}

\begin{table}
\begin{centering}
\begin{tabular}{|c|c|c|c|c||c|c|c|c|c|}
\hline
 &  & SR \cite{Kiselev2002vz}  & PM \cite{Kiselev2002vz}  & LF \cite{Wang2009xt}  &  &  & SR \cite{Kiselev2002vz}  & PM \cite{Kiselev2002vz}  & LF \cite{Wang2009xt}\tabularnewline
\hline
\hline
 & $F\left(0\right)$  & 1.3  & 1.1  & 0.73  &  & $F\left(0\right)$  & 8.1  & 8.2  & 6.1\tabularnewline
\hline
$F_{+}$  & $c_{1},\,\mathrm{GeV}^{-2}$  & 0.30  & 0.30  & 0.56  & $F_{0}^{A}$, GeV  & $c_{1},\,\mathrm{GeV}^{-2}$  & 0.30  & 0.52  & 0.56\tabularnewline
\hline
 & $c_{2},\,\mathrm{GeV}^{-4}$  & 0.069  & 0.069  & 0.030  &  & $c_{2},\,\mathrm{GeV}^{-4}$  & 0.069  & 0.02  & 0.087\tabularnewline
\hline
\hline
 & $F\left(0\right)$  & -5.8  & -5.9  & -1.7  &  & $F\left(0\right)$  & 0.15  & 0.30  & 0.30 \tabularnewline
\hline
$F_{-}$  & $c_{1},\,\mathrm{GeV}^{-2}$  & 0.30  & 0.30  & 0.70  & $F_{+}^{A}$, $\mathrm{GeV}^{-1}$  & $c_{1},\,\mathrm{GeV}^{-2}$  & 0.30  & 0.30  & 0.30 \tabularnewline
\hline
 & $c_{2},\,\mathrm{GeV}^{-4}$  & 0.069  & 0.069  & -0.02  &  & $c_{2},\,\mathrm{GeV}^{-4}$  & 0.069  & 0.069  & 0.069 \tabularnewline
\hline
\hline
 & $F\left(0\right)$, $\mathrm{GeV}^{-1}$  & 1.08  & 1.1  & 0.31  &  & $F\left(0\right)$  & 1.8  & 1.4  & 1.4 \tabularnewline
\hline
$F_{V}$  & $c_{1},\,\mathrm{GeV}^{-2}$  & 0.30  & 0.30  & 0.12  & $F_{-}^{A}$, $\mathrm{GeV}^{-1}$  & $c_{1},\,\mathrm{GeV}^{-2}$  & 0.30  & 0.30  & 0.30 \tabularnewline
\hline
 & $c_{2},\,\mathrm{GeV}^{-4}$  & 0.69  & 0.069  & -0.02  &  & $c_{2},\,\mathrm{GeV}^{-4}$  & 0.069  & 0.069  & 0.069 \tabularnewline
\hline
\end{tabular}
\par\end{centering}

\caption{Form-factor parameters\label{tab:FFS}}

\end{table}

\section{Numerical Results}

When inclusive decays $B_{c}\to B_{s}^{(*)}u\overline{d}$ are consideres,
corresponding spectral fiuctions are equal to\begin{eqnarray}
\rho_{L}^{ud}\left(q^{2}\right) & = & 0,\qquad\rho_{T}^{ud}\left(q^{2}\right)=\frac{1}{2\pi^{2}}.\label{eq:rhoQQ}\end{eqnarray}
Branching fracttions of these decays for listed above form-factor
sets are

\begin{eqnarray*}
\mathrm{Br}_{\mathrm{SR}}\left(B_{s}u\overline{d}\right) & = & \BrSbsSRqq\%,\,\,\,\mathrm{Br}_{\mathrm{\mathrm{PM}}}\left(B_{s}u\overline{d}\right)=\BrSbsPMqq\%,\,\,\,\mathrm{Br}_{\mathrm{LF}}\left(B_{s}u\overline{d}\right)=\BrSbsLFqq\%,\\
\mathrm{Br}_{\mathrm{SR}}\left(B_{s}^{*}u\overline{d}\right) & = & \BrVbsSRqq\%,\,\,\,\mathrm{Br}_{\mathrm{\mathrm{PM}}}\left(B_{s}^{*}u\overline{d}\right)=\BrVbsPMqq\%,\,\,\,\mathrm{Br}_{\mathrm{LF}}\left(B_{s}^{*}u\overline{d}\right)=\BrVbsLFqq\%.\end{eqnarray*}
The braching fractions of semileptonic decays $B_{c}\to B_{s}^{(*)}e\nu_{e}$
can be obtained from these values by simple substitution\begin{eqnarray*}
\rho_{L}^{e\nu}\left(q^{2}\right) & = & 0,\qquad\rho_{T}^{e\nu}\left(q^{2}\right)=\frac{1}{N_{c}a_{1}^{2}}\rho_{T}^{ud}\left(q^{2}\right).\end{eqnarray*}

Let us now consider $B_{c}\to B_{s}^{(*)}\pi^{+}$ decay. The $W\to\pi$
vertex is written as

\begin{eqnarray}
\left\langle \pi^{+}\left|J_{\mu}\right|W\right\rangle  & = & f_{\pi}q_{\mu},\label{eq:pi}\end{eqnarray}
 where $q_{\mu}$ is $\pi$-meson momentum and constant $f_{\pi}$
can be determined, for example, from the width of leptonic decay $\pi^{+}\to e^{+}\nu+e$.
Using experimental value for the width of this decay we obtain $f_{\pi}\approx130$
MeV. Spectral functions, that correspond to vertex (\ref{eq:pi})
are

\begin{eqnarray*}
\rho_{L}^{\pi}\left(q^{2}\right) & = & f_{\pi}^{2}\delta\left(q^{2}-m_{\pi}^{2}\right),\qquad\rho_{T}^{\pi}\left(q^{2}\right)=0.\end{eqnarray*}
 The values of $B_{c}\to B_{s}^{(*)}\pi$ branching fractions for
listed in the previous section form-factor sets are

\begin{eqnarray*}
\mathrm{Br}_{\mathrm{SR}}\left(B_{s}\pi\right) & = & \BrSbsSRpi\%,\,\,\,\mathrm{Br}_{\mathrm{\mathrm{PM}}}\left(B_{s}\pi\right)=\BrSbsPMpi\%,\,\,\,\mathrm{Br}_{\mathrm{LF}}\left(B_{s}\pi\right)=\BrSbsLFpi\%,\\
\mathrm{Br}_{\mathrm{SR}}\left(B_{s}^{*}\pi\right) & = & \BrVbsSRpi\%,\,\,\,\mathrm{Br}_{\mathrm{\mathrm{PM}}}\left(B_{s}^{*}\pi\right)=\BrVbsPMpi\%,\,\,\,\mathrm{Br}_{\mathrm{LF}}\left(B_{s}^{*}\pi\right)=\BrVbsLFpi\%.\end{eqnarray*}
It is clearly seen, that the difference in form-factors leads difference
in the branching fractions of these decays. All these branching fractions,
in the other hand, are about an order of magnitude higher, that the
branching fractions of $B_{c}\to J/\psi\pi$ decay \cite{Likhoded2009}.
The reason is that on the quark level $B_{c}\to B_{s}^{(*)}+\mathcal{R}$
decays are caused by Cabbibo-allowed $c\to su\bar{d}$ decay, while
$B_{c}\to J/\psi+\mathcal{R}$ decays are Cabbibo-suppressed.

If there are two $\pi$-mesons in the final state, the main decay
mode would be $B_{c}\to B_{s}^{(*)}\rho^{+}\to B_{s}^{(*)}\pi^{+}\pi^{0}$.
The vertex of $\rho$-meson interaction with $W$-boson has the form

\begin{eqnarray*}
\left\langle \rho^{+}\left|J_{\mu}\right|W\right\rangle  & = & f_{\rho}m_{\rho}\epsilon_{\mu},\end{eqnarray*}
 where $m_{\rho}$ and $\epsilon_{\mu}$are $\rho$-meson mass and
polarization vector, while $f_{\rho}$-constant is $f_{\rho}\approx210$
MeV. In the limit of zero $\rho$-meson widths the spectral functions
$\rho_{T,L}^{\rho}\left(q^{2}\right)$ are equal to

\begin{eqnarray*}
\rho_{L}^{\rho}\left(q^{2}\right) & = & 0,\qquad\rho_{T}^{\rho}\left(q^{2}\right)=f_{\rho}^{2}\delta\left(q^{2}-m_{\rho}^{2}\right).\end{eqnarray*}
 These spectral functions lead to following branching fractions of
$B_{c}\to B_{s}^{(*)}\rho$ decays:

\begin{eqnarray}
\mathrm{Br}_{\mathrm{SR}}\left(B_{s}\rho\right) & = & \BrSbsSRrho\%,\,\,\,\mathrm{Br}_{\mathrm{\mathrm{PM}}}\left(B_{s}\rho\right)=\BrSbsPMrho\%,\,\,\,\mathrm{Br}_{\mathrm{LF}}\left(B_{s}\rho\right)=\BrSbsLFrho\%,\label{eq:BrSrho}\\
\mathrm{Br}_{\mathrm{SR}}\left(B_{s}^{*}\rho\right) & = & \BrVbsSRrho\%,\,\,\,\mathrm{Br}_{\mathrm{\mathrm{PM}}}\left(B_{s}^{*}\rho\right)=\BrVbsPMrho\%,\,\,\,\mathrm{Br}_{\mathrm{LF}}\left(B_{s}^{*}\rho\right)=\BrVbsLFrho\%.\label{eq:BrVrho}\end{eqnarray}
 As in previous case, these values exceed significantly the branching
fractions of $B_{c}\to J/\psi\rho$ decay.

It should be noted, however, that in contrast to $B_{c}\to J/\psi+2\pi$
decay, for $B_{c}\to B_{s}^{(*)}+2\pi$ decays it is not valid to
neglect the width of $\rho$-meson. Because of the small mass difference
$M_{B_{c}}-M_{B_{s}}\approx1$ GeV $\rho$-meson is almost at the
end of the phase space, so its non-zero width changes significantly
$q^{2}$-distributions for these decays. For this reason one should
use more realistic parametrization for spectral functions $\rho_{L,T}^{2\pi}\left(q^{2}\right)$.
Due to vector current conservation the longitudinal spectral function
$\rho_{L}^{2\pi}\left(q^{2}\right)=0$. The information on transverse
spectral function can be obtained from analysis of $\tau$-lepton
decay $\tau\to\nu_{\tau}+2\pi$. For this decay $q^{2}$-distribution
has the form

\begin{eqnarray}
\frac{d\Gamma\left(\tau\to\nu_{\tau}\mathcal{R}\right)}{dq^{2}} & = & \frac{G_{F}^{2}}{16\pi m_{\tau}}\frac{\left(m_{\tau}^{2}-q^{2}\right)^{2}}{m_{\tau}^{3}}\left(m_{\tau}^{2}+2q^{2}\right)\rho_{T}^{\mathcal{R}}\left(q^{2}\right),\label{eq:tau}\end{eqnarray}
 where $m_{\tau}$ is $\tau$-lepton mass, and, in the framework of
factorization theorem, spectral function $\rho_{T}^{2\pi}\left(q^{2}\right)$
is universal, that is independent on $\pi$-pair production dynamics.
The authors of paper \cite{Kuhn1990} give the parametrization for
this spectral function with $\rho$-, $\rho'$- and $\omega$-meson
contributions taken into account. In our article we use a more simple
parametrization

\begin{eqnarray*}
\rho_{T}^{2\pi}\left(s\right) & = & 1.35\times10^{-3}\left(\frac{s-4m_{\pi}^{2}}{2}\right)^{2}\frac{1+0.64s}{\left(s-0.57\right)^{2}+0.013}\end{eqnarray*}
 where $s$ is measured in $\mathrm{GeV}^{2}$ (the spectral function
itself is dimensionless)

\begin{figure}
\begin{centering}
\includegraphics[width=17cm]{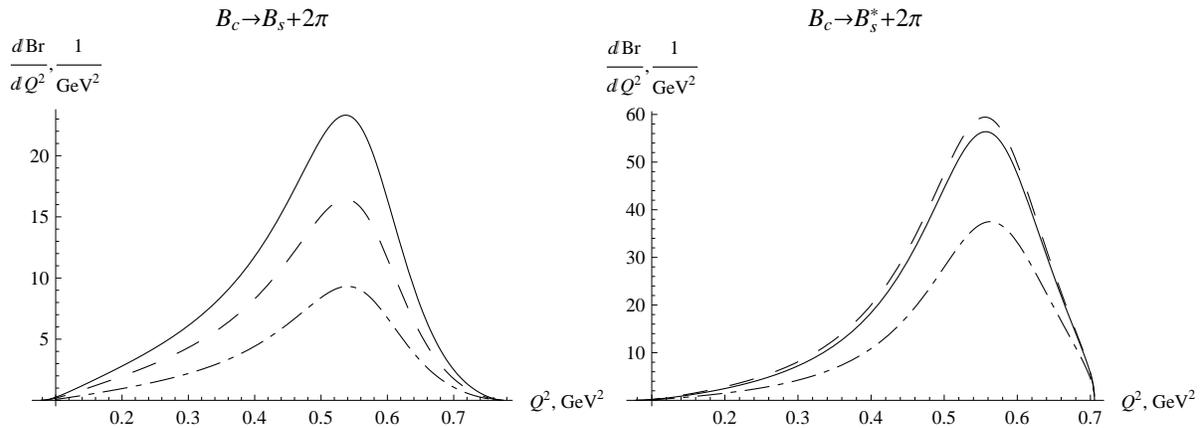}
\par\end{centering}

\caption{Distributions over $q^{2}$ of the branching fractions of $B_{c}\to B_{s}+2\pi$
decay (left figure) and $B_{c}\to B_{s}^{*}+2\pi$ decay (right panel).
Solid, dashed and dash-dotted lined in these figure correspond to
{}``SR'', {}``PM'' and {}``LF'' form-factor sets respectively\label{fig:dG2}}

\end{figure}

\begin{figure}
\begin{centering}
\includegraphics[width=17cm]{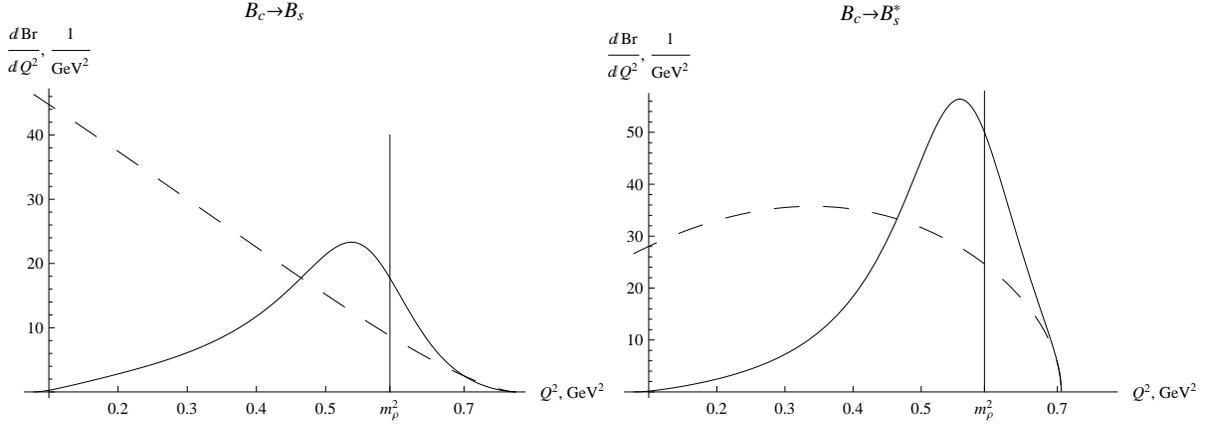}
\par\end{centering}

\caption{Distributions of branching fractions of $B_{c}\to B_{s}+2\pi$, $B_{c}\to B_{s}+u\bar{d}$
decays (left panel) and $B_{c}\to B_{s}^{*}+2\pi$, $B_{c}\to B_{s}^{*}+u\bar{d}$
(right panel) over the squared transferred momentum. On both figures
solid and dashed lines line correspond to $\pi\pi$ and $u\bar{d}$
final states respectively. Experimental value of $\rho$-meson mass
is shown by vertical lines.\label{fig:dG2pi}}

\end{figure}

Distributions of $B_{c}\to B_{s}^{(*)}+2\pi$-decays branching fractions
over squared transferred momentum $q^{2}$ for listed in the previous
section form-factor sets are presented in fig.\ref{fig:dG2}. Solid,
dashed and dash-dotted lines in this figure correspond to form-factor
sets {}``SR'', {}``QM'' and {}``LF'' respectively. Corresponding
branching fractions are\begin{eqnarray}
\mathrm{Br}_{\mathrm{SR}}\left(B_{s}\pi\pi\right) & = & \BrSbsSRpp\%,\,\mathrm{Br}_{\mathrm{PM}}\left(B_{s}\pi\pi\right)=\BrSbsPMpp\%,\,\mathrm{Br}_{\mathrm{LF}}\left(B_{s}\pi\pi\right)=\BrSbsLFpp\%,\label{eq:brS2pi}\\
\mathrm{Br}_{\mathrm{SR}}\left(B_{s}^{*}\pi\pi\right) & = & \BrVbsSRpp\%,\,\mathrm{Br}_{\mathrm{PM}}\left(B_{s}^{*}\pi\pi\right)=\BrVbsPMpp\%,\,\mathrm{Br}_{\mathrm{LF}}\left(B_{s}^{*}\pi\pi\right)=\BrVbsLFpp\%.\label{eq:brV2pi}\end{eqnarray}
One can easily see that these branching fractions are smaller than
presented above values (\ref{eq:BrSrho}), (\ref{eq:BrVrho}). In
fig.\ref{fig:dG2pi} we present $q^{2}$-distributions of $B_{c}\to B_{s}^{(*)}+2\pi$
decay (solid line) and $B_{c}\to B_{s}^{(*)}+u\bar{d}$ (dashed line.
In this case the spectral function does not depend on $q^{2}$, see
eq.(\ref{eq:rhoQQ}) ). Experimental value of $\rho$-meson mass is
shown on this figure by vertical lines. From this figure it is clear,
that in the region $q^{2}\approx\left(m_{\rho}\pm\Gamma_{\rho}/2\right)^{2}$,
where spectral function is significantly non-zero, the distributions
$d\Gamma\left(B_{c}\to B_{s}^{(*)}+u\bar{d}\right)/dq^{2}$ vary strongly,
so one cannot neglect $\rho$-meson width in this case. This is the
reason for large difference between (\ref{eq:BrSrho}), (\ref{eq:BrVrho})
and (\ref{eq:brS2pi}), (\ref{eq:brV2pi}) branching fractions.

\begin{figure}
\begin{centering}
\includegraphics[width=17cm]{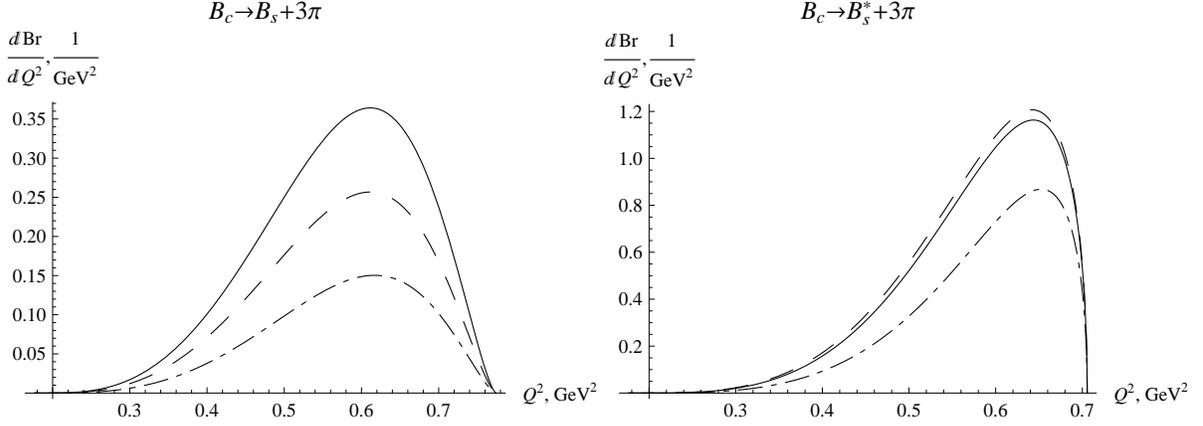}
\par\end{centering}

\caption{Distributions of $B_{c}\to B_{s}^{(*)}+3\pi$ branching fractions
over squared momentum of $3\pi$-meson system. Notations are same
as in fig.\ref{fig:dG2}.\label{fig:dG3}}

\end{figure}

If there are three $\pi$-mesons in the final state the main production
mode would be $a_{1}\to\rho\pi\to3\pi$. There are two possible charge
configurations ($\pi^{+}\pi^{-}\pi^{+}$ and $\pi^{+}\pi^{0}\pi^{0}$),
and in our article $3\pi$ stands for sum of these final states. Because
of partial conservation of axial current longitudinal spectral function
$\rho_{L}^{3\pi}(q^{2})$ could be set equal to zero. In paper \cite{Kuhn1990}
the parametrization of transverse spectral function $\rho_{T}^{4\pi}\left(q^{2}\right)$,
expressed through mass and width of $a_{1}$-meson is proposed. In
our article we use a simpler parametrisation, obtained from fit of
experimental $q^{2}$ distribution of $\tau\to\nu_{\tau}+3\pi$ decay:\begin{eqnarray*}
\rho_{T}^{3\pi}\left(s\right) & = & 5.86\times10^{-5}\left(\frac{s-9m_{\pi}^{2}}{s}\right)^{4}\frac{1+190s}{\left[\left(s-1.06\right)^{2}+0.48\right]^{2}},\end{eqnarray*}
 where $s$ is measured in $\mathrm{GeV}^{2}$. Distributions of $B_{c}\to B_{s}^{(*)}+3\pi$
branching fractions over $q^{2}$ for different stets of form-factors
are shown in fig.\ref{fig:dG3}. Corresponding branching fractions
are equal to \begin{eqnarray*}
\mathrm{Br}_{\mathrm{SR}}\left(B_{s}3\pi\right) & = & \BrSbsSRppp\%,\,\mathrm{Br}_{\mathrm{PM}}\left(B_{s}3\pi\right)=\BrSbsPMppp\%,\,\mathrm{Br}_{\mathrm{LF}}\left(B_{s}3\pi\right)=\BrSbsLFppp\%,\\
\mathrm{Br}_{\mathrm{SR}}\left(B_{s}^{*}3\pi\right) & = & \BrVbsSRppp\%,\,\mathrm{Br}_{\mathrm{PM}}\left(B_{s}^{*}3\pi\right)=\BrVbsPMppp\%,\,\mathrm{Br}_{\mathrm{LF}}\left(B_{s}^{*}3\pi\right)=\BrVbsLFppp\%.\end{eqnarray*}
 One can clearly seen, that these values are significantly smaller,
than presented above branching fractions of the decays $B_{c}\to B_{s}^{(*)}+\pi$,
$B_{c}\to B_{s}^{(*)}+2\pi$. This decays, however, could be interesting
from experimental point of view, since in $\pi^{+}\pi^{-}\pi^{+}$
charge configuration $\pi^{0}$-meson, whose registration could be
problematic, is absent. Moreover, in contrast to $B_{c}\to B_{s}^{(*)}+\pi$,
$B_{c}\to B_{s}^{(*)}+2\pi$ decays, the branching fractions of $B_{c}\to B_{s}^{(*)}+3\pi$
decay are smaller than the corresponding branching fractions of $B_{c}\to J/\psi+3\pi$
decay. The reason is that for $B_{s}^{(*)}$-meson in the final sate
the masses of initial and final mesons are rather close, so phase-space
suppression compensates the enhancement caused by CKM matrix element.

\begin{figure}
\begin{centering}
\includegraphics[width=17cm]{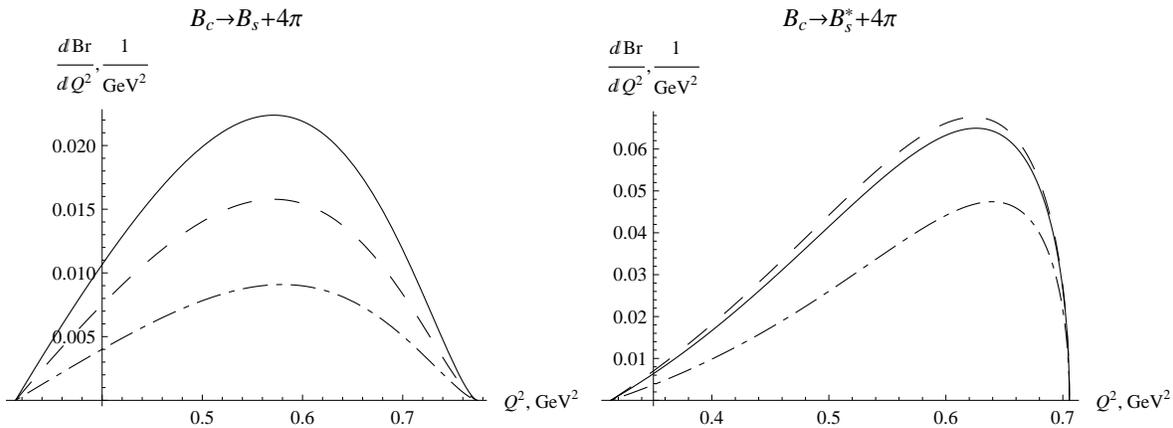}
\par\end{centering}

\caption{Distributions of $B_{c}\to B_{s}^{(*)}+4\pi$ branching fractions
over squared momentum of $4\pi$-meson system. Notations are same
as in fig.\ref{fig:dG2}, \ref{fig:dG3}.\label{fig:dG4}}

\end{figure}

If there are four $\pi$-mesons in the final state two charge configurations
are possible: $\pi^{+}\pi^{-}\pi^{+}\pi^{0}$ and $\pi^{+}\pi^{0}\pi^{0}\pi^{0}$,
in what follows $4\pi$ stands for sum of these configurations. Due
to vector colour conservation the longitudinal spectral function $\rho_{L}^{4\pi}(q^{2})=0$.
Information about traverse spectral function $\rho_{T}^{4\pi}\left(q^{2}\right)$
is available from experimental data on $\tau\to\nu_{\tau}+4\pi$ decay
(see eq.(\ref{eq:tau}) ) or electron-positron annihilation $e^{+}e^{-}\to4\pi$.
Cross section of the latter reaction is\begin{eqnarray*}
\sigma\left(e^{+}e^{-}\to4\pi\right) & = & \frac{4\pi\alpha^{2}}{s}\rho_{T}^{4\pi}(s).\end{eqnarray*}
 In both cases we have similar from of the spectral function, that
can be parametrized by the expression

\begin{eqnarray*}
\rho_{T}^{4\pi}\left(s\right) & \approx & 1.8\times10^{-4}\left(\frac{s-16m_{\pi}^{2}}{s}\right)\frac{1-5.07s+8.63s^{2}}{\left[\left(s-1.83\right)^{2}+0.61\right]^{2}}.\end{eqnarray*}
In fig.\ref{fig:dG4} $q^{2}$-distributions of $B_{c}\to B_{s}^{(*)}+4\pi$
decay branching fractions for different form-factor sets are shown.
Numerical values of these branching fractions are\begin{eqnarray*}
\mathrm{Br}_{\mathrm{SR}}\left(B_{s}4\pi\right) & = & \BrSbsSRpppp\%,\,\mathrm{Br}_{\mathrm{PM}}\left(B_{s}4\pi\right)=\BrSbsPMpppp\%,\,\mathrm{Br}_{\mathrm{LF}}\left(B_{s}4\pi\right)=\BrSbsLFpppp\%,\\
\mathrm{Br}_{\mathrm{SR}}\left(B_{s}^{*}4\pi\right) & = & \BrVbsSRpppp\%,\,\mathrm{Br}_{\mathrm{PM}}\left(B_{s}^{*}4\pi\right)=\BrVbsPMpppp\%,\,\mathrm{Br}_{\mathrm{LF}}\left(B_{s}^{*}4\pi\right)=\BrVbsLFpppp\%.\end{eqnarray*}
 These values are significantly smaller then presented above branching
fractions of one, two, or three $\pi$-meson production, as well as
branching fractions of $B_{c}\to J/\psi+4\pi$ decay \cite{Likhoded2009}.
The reason is mentioned above phase-space suppression in $B_{c}\to B^{(*)}+4\pi$
reaction. Thus we can see, that for all form-factor sets dominant
decay mode $c\to s$ is saturated by $B_{c}\to B_{s}^{(*)}\pi$, $\rho$
decays.

In table \ref{tab:Br} branching obtained in our article fractions
of $B_{c}\to B_{s}^{(*)}+\mathcal{R}$ decays for different form-factor
sets are listed. It is interesting to compare these results with simple
estimates based on duality relations. According to paper \cite{Gershtein1995}
$B_{c}$-meson decays with spectator $b$-quark takes about $45\%$
of total $B_{c}$-meson width. These decays are almost saturated by
$B_{c}\to B_{s}^{(*)}+X$ decays, so the sum of $B_{c}\to B_{s}^{(*)}+n\pi$
branching fractions should be about this value. It can be easily seen,
that this is true for {}``SR'' and {}``PM'' form-factor sets (the
corresponding sum for them is 44\% and 40\% respectively). For {}``LF''
form-factors set this sum is significantly smaller ($\sim23\%$).
From table \ref{tab:Br} it can be seen also, that for every decay
mode the branching fraction obtained with {}``LF'' form-factors
sets is smaller then those obtained with {}``SM'' and {}``PM''
form-factors sets. As a result, the total width of $B_{c}\to B_{s}^{(*)}+X$
inclusive decay (and, hence, lifetime of $B_{c}$-meson) would contradict
experimental value.

\begin{table}
\begin{centering}
\begin{tabular}{|c|c|c|c||c|c|c|c|}
\hline
\multicolumn{4}{|c||}{$\mathrm{Br\left(B_{c}\to B_{s}+\mathcal{R}\right)},\,\%$} & \multicolumn{4}{c|}{$\mathrm{Br\left(B_{c}\to B_{s}^{*}+\mathcal{R}\right)},\,\%$}\tabularnewline
\hline
$\mathcal{R}$  & SR \cite{Kiselev2002vz}  & PM \cite{Kiselev2002vz}  & LF \cite{Wang2009xt}  & $\mathcal{R}$  & SR \cite{Kiselev2002vz}  & PM \cite{Kiselev2002vz}  & LF \cite{Wang2009xt} \tabularnewline
\hline
\hline
$\pi$  & $\BrSbsSRpi$  & $\BrSbsPMpi$  & $\BrSbsLFpi$  & $\pi$  & $\BrVbsSRpi$  & $\BrVbsPMpi$  & $\BrVbsLFpi$\tabularnewline
\hline
$\rho$  & $\BrSbsSRrho$  & $\BrSbsPMrho$  & $\BrSbsLFrho$  & $\rho$  & $\BrVbsSRrho$  & $\BrVbsPMrho$  & $\BrVbsLFrho$ \tabularnewline
\hline
$2\pi$  & $\BrSbsSRpp$  & $\BrSbsPMpp$  & $\BrSbsLFpp$  & $2\pi$  & $\BrVbsSRpp$  & $\BrVbsPMpp$  & $\BrVbsLFpp$\tabularnewline
\hline
$3\pi$  & $\BrSbsSRppp$  & $\BrSbsPMppp$  & $\BrSbsLFppp$  & $3\pi$  & $\BrVbsSRppp$  & $\BrVbsPMppp$  & $\BrVbsLFppp$\tabularnewline
\hline
$4\pi$  & $\BrSbsSRpppp$  & $\BrSbsPMpppp$  & $\BrSbsLFpppp$  & $4\pi$  & $\BrVbsSRpppp$  & $\BrVbsPMpppp$  & $\BrVbsLFpppp$\tabularnewline
\hline
\hline
$u\bar{d}$  & $\BrSbsSRqq$  & $\BrSbsPMqq$  & $\BrSbsLFqq$  & $u\bar{d}$  & $\BrVbsSRqq$  & $\BrVbsPMqq$  & $\BrVbsLFqq$ \tabularnewline
\hline
\end{tabular}
\par\end{centering}

\caption{Branching fractions of $B_{c}\to B_{s}^{(*)}+\mathcal{R}$ decays\label{tab:Br}}

\end{table}

\section{Conclusion}

The article is devoted to study of exclusive $B_{c}$-meson decays
with production of $B_{s}^{(*)}$-meson an light hadron system $n\pi$
with $n=1$, $2$, $3$ or $4$.

$B_{c}$-mesons, that is particles that in valence approximation are
build from $c$- and $b$-quarks, take the intermediate place between
charmonia ($\bar{c}c$-mesons) and bottomonia ($\bar{b}b$-mesons),
so they can be used for independent check of theoretical models used
for analysis of heavy quarkonia with hidden flavour. Only ground state
of $B_{c}$-meson family is now obsevrved experimentally, with only
mass and lifetime measured with good accuracy. One can expect large
$B_{c}$-meson production at LHC collider, so branching fractions
of exclusive $B_{c}$-meson decays could be measured and theoretical
prediction are desirable.

Previous works were mainly devoted to two-particle decays of $B_{c}$-meson
(see, for example, \cite{Gershtein1995,Kiselev2000nf,Kiselev2000pp,Kiselev2001zb,Hernandez2006gt,Huang2008zg,Sun2008wa,Sun2008ew}).
In our recent article \cite{Likhoded2009} we, on the contrary, consider
decays $B_{c}\to J/\psi+n\pi$ with $n=1$, $2$, $3$ or $4$. In
the framework of factorization theorem the branching fractions of
these decays are written as convolution of hard part, describing $B_{c}\to J/\psi W$
decay and spectral functions, responsible for $W\to n\pi$ transition.
Form factors of $B_{c}\to J/\psi W$ vertex can be determined from
different theoretical models (QCD sum rules, potential models, light-front
covariant quark models, etc), spectral functions --- from analysis
of $\tau$-lepton decays and electron-positron annihilation.

In the present article we consider in the same approach the decays
$B_{c}\to B_{s}^{(*)}+n\pi$ with $n=1,\dots,4$. Using different
sets of $B_{c}\to B_{s}^{(*)}W$ vertex form-factors we calculated
branching fractions of these decays and distributions over the squared
momentum of $\pi$-mesons system. In contrast to $B_{c}\to J/\psi+n\pi$
decays these reactions are Cabbibo-allowed, so for one and two $\pi$-mesons
in the final state their branching fractions are greater, than the
branching fractions of corresponding $B_{c}\to J/\psi+n\pi$ decays.
If the number of $\pi$-mesons in the final state is larger, the suppression
caused by small phase-space in the reaction $B_{c}\to B_{s}^{(*)}+X$
is important and $B_{c}\to J/\psi+X$ decays dominate.

We would like to note, that considered in our article decays are suitable
for $B_{c}$-meson study at hadron colliders. Besides large branching
fractions there are also small background processes to these decays.

This work was supported in part by Russian foundation for basic research
(grant \#10-02-00061a). One of the authors (A.V.L.) was also supported
by Russian Foundation of Science Support, non-commercial foundations
{}``Dynasty'' and the grant of the president of Russian Federation
(grant \#MK-406.2010.2).


\end{document}